\documentclass[cameraready]{Interspeech}

\usepackage{comment}
\usepackage{amsmath}
\usepackage{amssymb}
\usepackage{booktabs}
\usepackage{multirow}
\usepackage{graphicx}
\usepackage{url}
\usepackage{subfigure}
\usepackage{amsfonts}
\usepackage{pifont}
\newcommand{\cmark}{\ding{51}}  
\newcommand{\xmark}{\ding{55}}  

\usepackage{tcolorbox}
\usepackage[T1]{fontenc}
\tcbuselibrary{skins,breakable}

\usepackage{colortbl}
\definecolor{highlight}{HTML}{E2F0D9} 

\title{Audio-DeepThinker: Progressive Reasoning-Aware Reinforcement Learning for High-Quality Chain-of-Thought Emergence in Audio Language Models}

\author[affiliation={1}, equalcontribution]{Xiang}{He}
\author[affiliation={1}, equalcontribution]{Chenxing}{Li}
\author[affiliation={2}]{Jinting}{Wang}
\author[affiliation={2}]{Yan}{Rong}
\author[affiliation={2}]{Tianxin}{Xie}
\author[affiliation={1}]{Wenfu}{Wang}
\author[affiliation={2}, orcid=0000-0002-4497-0135, correspondingauthor]{Li}{Liu}
\author[affiliation={1}, orcid=0000-0003-0520-6844, correspondingauthor]{Dong}{Yu}

\address{
    $^1$ Tencent AI Lab, Beijing, China \\
    $^2$ The Hong Kong University of Science and Technology (Guangzhou), Guangzhou, China
}

\email{avrillliu@hkust-gz.edu.cn, dyu@global.tencent.com}


\begin{document}

\maketitle

\begin{abstract}
Large Audio-Language Models (LALMs) have made significant progress in audio understanding, yet they primarily operate as perception-and-answer systems without explicit reasoning processes. Existing methods for enhancing audio reasoning rely either on supervised chain-of-thought (CoT) fine-tuning, which is limited by training data quality, or on reinforcement learning (RL) with coarse rewards that do not directly evaluate reasoning quality. As a result, the generated reasoning chains often appear well-structured yet lack specific acoustic grounding.
We propose Audio-DeepThinker, a framework built on two core ideas. First, we introduce a hybrid reasoning similarity reward that directly supervises the quality of generated reasoning chains by combining an LLM evaluator assessing logical path alignment, key step coverage, and analytical depth with an embedding similarity component enforcing semantic alignment with reference reasoning chains. Second, we propose a progressive two-stage curriculum that enables high-quality CoT reasoning to emerge through pure RL exploration, without any supervised reasoning fine-tuning, from an instruction-tuned model that possesses no prior chain-of-thought capability. Stage~1 trains on foundational audio QA with the hybrid reward to foster basic reasoning patterns, while Stage~2 shifts to acoustically challenging boundary cases with an LLM-only reward for greater reasoning diversity. Audio-DeepThinker achieves state-of-the-art results on MMAR (74.0\%), MMAU-test-mini (78.5\%), and MMSU (77.26\%), winning 1st Place in the Interspeech 2026 Audio Reasoning Challenge (Single Model Track). Interpretability analyses further reveal that RL training primarily reshapes upper-layer MoE gating mechanisms and that reasoning tokens crystallize progressively in the upper transformer layers, offering mechanistic insights into how audio reasoning emerges through exploration.
\end{abstract}

\section{Introduction}
\label{sec:intro}

Recent progress in large language models has demonstrated that reinforcement learning (RL) can unlock sophisticated reasoning capabilities through pure exploration, as exemplified by DeepSeek-R1~\cite{guo2025deepseek} and OpenAI o1~\cite{jaech2024openai}. These successes have inspired growing interest in extending RL-based reasoning to multimodal domains~\cite{huang2025vision, liu2025visual, xu2025qwen3}. In the audio domain, Large Audio-Language Models (LALMs) such as Qwen2-Audio~\cite{chu2024qwen2}, SALMONN~\cite{tang2023salmonn}, and Step-Audio 2~\cite{wu2025step} have made significant strides in audio understanding, yet they primarily operate as perception-and-answer systems without explicit reasoning processes.

\begin{figure*}[t]
    \centering
    \includegraphics[width=1.\textwidth]{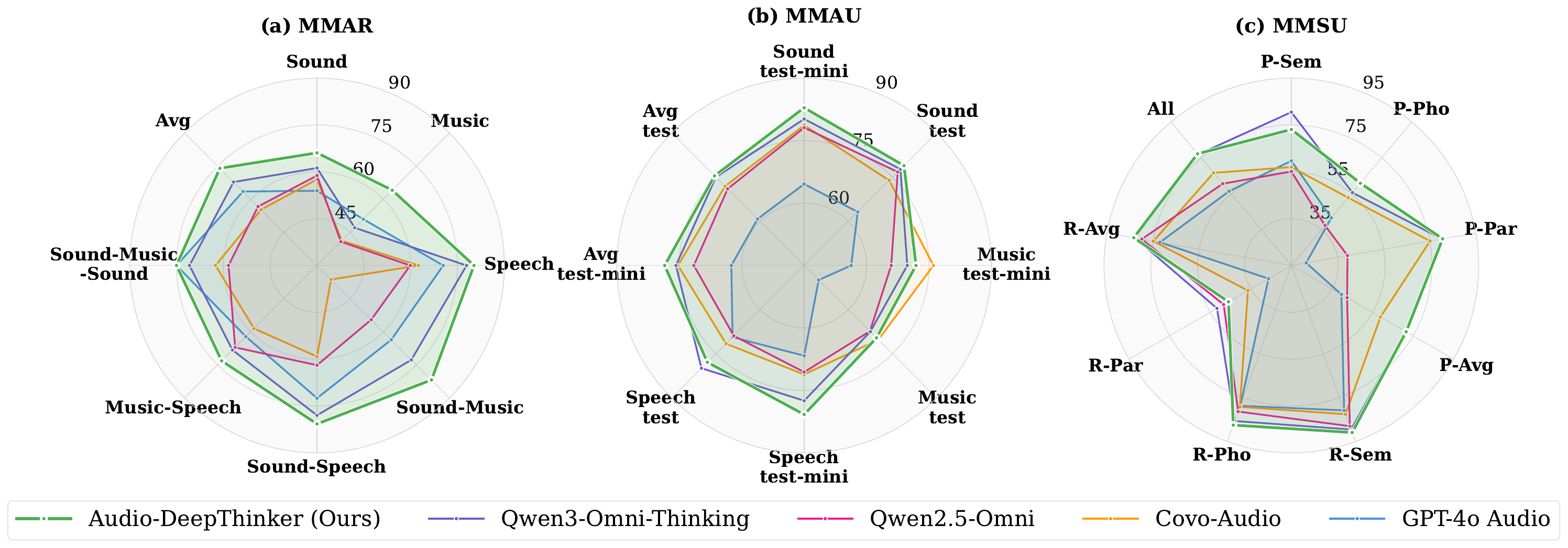}
    \caption{Radar chart comparison of Audio-DeepThinker against representative open-source and closed-source models on three benchmarks: (a) MMAR, (b) MMAU, and (c) MMSU. Our method (green) achieves consistently strong performance across all three benchmarks.}
    \label{fig:performance}
\end{figure*}

To bridge the gap between perception and reasoning, prior works have explored two paradigms. Supervised approaches such as Audio Flamingo 2~\cite{ghosh2025audio} and Audio Flamingo 3~\cite{goel2025audio} construct curated Chain-of-Thought (CoT) annotations and fine-tune models to imitate them. While effective, these methods are fundamentally limited by the quality and diversity of human-authored reasoning, and cannot discover novel reasoning strategies beyond the training data. Reinforcement Learning (RL)-based approaches offer a more flexible alternative. Early methods such as R1-AQA~\cite{li2025reinforcement}, Omni-R1~\cite{rouditchenko2025omni}, and AudioMCQ~\cite{he2025measuring} apply GRPO~\cite{shao2024deepseekmath} with accuracy and format rewards, demonstrating that RL can improve audio QA performance. More recent works have begun to incorporate reasoning-related signals. For example, Audio-Thinker~\cite{wu2025audio} introduces adaptive rewards to guide when the model should reason, and CESAR~\cite{fan2025incentivizing} introduces a comprehensive suite to reward structured patterns and causal logic.

Despite these advances, two fundamental challenges remain: (i) how to ensure that the generated reasoning chains are genuinely grounded in the audio content, rather than being formally well-formatted yet semantically decoupled from the actual acoustic evidence; and (ii) how to fundamentally improve the quality of the reasoning process, ensuring that generated reasoning chains are logically coherent, well-structured, and free from misleading or spurious inference steps. While recent methods such as Audio-Thinker~\cite{wu2025audio} and CESAR~\cite{fan2025incentivizing} have made initial strides toward reasoning supervision, they do not assess the content of the reasoning chain at a fine-grained level: whether the intermediate steps are grounded in specific acoustic evidence and whether the logical path is analytically sound. This limitation arises because the reasoning chains are generated solely by the model itself, without any ground-truth reference chains to serve as a basis for evaluating their quality.

To this end, we propose Audio-DeepThinker, a framework that addresses both challenges through two complementary innovations. \textbf{First}, we develop a hybrid reasoning similarity reward that provides direct, fine-grained supervision over both the audio grounding and logical quality of generated reasoning. This reward combines an LLM evaluator assessing logical path alignment, key step coverage, and analytical depth with embedding similarity enforcing semantic alignment against reference reasoning chains. Since the reference chains are derived from detailed audio captions, aligning with them inherently encourages the model to ground its reasoning in acoustic content. Critically, this reward is applied only when the answer is correct, shifting the objective from ``finding any path to the answer'' to ``constructing a logically rigorous and faithfully grounded reasoning chain.'' \textbf{Second}, to fully leverage this reward signal, we propose a progressive two-stage RL curriculum that enables high-quality CoT to emerge via pure RL exploration from an instruction-tuned model without prior reasoning capability. Stage~1 fosters basic reasoning patterns on foundational audio QA with comprehensive reward supervision, while Stage~2 refines these on acoustically challenging boundary cases with a streamlined reward encouraging diverse reasoning strategies. The emergence of reasoning suggests that the bottleneck in audio reasoning lies not in model architecture, but in the lack of reward signals providing fine-grained reasoning supervision.

Our main contributions are summarized as follows:
\begin{itemize}
    \item We propose a hybrid reasoning similarity reward that directly supervises the quality of generated reasoning chains by combining LLM-based logical evaluation with embedding-based semantic alignment, jointly addressing the dual challenges of reasoning soundness and audio grounding that existing RL methods leave unresolved.

    \item We demonstrate that high-quality CoT reasoning can emerge through pure RL exploration from an instruction-tuned LALM that has no prior chain-of-thought capability, without any supervised reasoning fine-tuning. This is enabled by a progressive two-stage curriculum: Stage~1 establishes foundational reasoning patterns with comprehensive reward supervision, while Stage~2 enhances performance on challenging boundary cases with a streamlined, exploration-friendly reward.

    \item Audio-DeepThinker achieves state-of-the-art results on MMAR (74.0\%), MMAU-test-mini (78.5\%), and MMSU (77.26\%), as visualized in Fig.~\ref{fig:performance}, winning 1st Place in the Single Model Track of the Interspeech 2026 Audio Reasoning Challenge.

    \item Mechanistic interpretability analyses reveal that RL training primarily reshapes upper-layer MoE gating mechanisms and that reasoning-related tokens crystallize progressively in the upper layers (L40+) before answer production, providing novel insights into how audio reasoning capabilities are acquired through exploration.
\end{itemize}

\section{Related Work}
\label{sec:related}

\noindent\textbf{Large Audio-Language Models (LALMs).}
Significant progress has been made in audio understanding through large-scale pre-training. SALMONN~\cite{tang2023salmonn} proposed a generic hearing framework for LLMs. Audio Flamingo~\cite{kong2024audio} introduced few-shot capabilities and dialogue abilities, while Qwen2-Audio-Instruct~\cite{chu2024qwen2} pushed the performance of open-source models through large-scale audio-text pre-training. These were followed by GPT-4o and GPT-4o mini Audio~\cite{hurst2024gpt}, which natively integrated multimodal reasoning, and Gemini 2.0 Flash~\cite{universalaiassistant}, which demonstrated strong results across universal assistance tasks. More recently, Baichuan-Omni-1.5~\cite{li2025baichuan}, Qwen-2.5-Omni~\cite{xu2025qwen25omnitechnicalreport}, and Step-Audio 2~\cite{wu2025step} established new baselines for full-modal interaction. However, these models primarily focus on direct perception, often lacking the explicit reasoning chains necessary for complex acoustic inference.

\noindent\textbf{Audio Reasoning with Supervised Learning.}
To bridge the gap between perception and reasoning, several works have explored supervised paradigms. Audio Flamingo 2~\cite{ghosh2025audio} utilized synthetic data and curriculum learning to enhance expert-level reasoning, while Audio Flamingo 3~\cite{goel2025audio} introduced the AF-Think dataset to enable on-demand thinking prefixes. While effective, these supervised approaches are constrained by the quality of curated annotations and do not allow the model to explore optimal reasoning paths autonomously.

\noindent\textbf{Reinforcement Learning for Audio Reasoning.}
Several methods have applied reinforcement learning to bridge the gap between perception and complex inference. Early attempts such as Audio-Reasoner~\cite{xie2025audio} and R1-AQA~\cite{li2025reinforcement} demonstrated that RL can outperform supervised fine-tuning in audio question answering. These were followed by Ke-Omni-R~\cite{zhao2025keomnir}, Omni-R1~\cite{rouditchenko2025omni}, and AudioMCQ~\cite{he2025measuring}, which utilized GRPO to optimize answer accuracy and format. However, these approaches often leave the intermediate reasoning process unconstrained. To address this, Audio-Thinker~\cite{wu2025audio} introduced rewards to guide when the model should think. Most recently, CESAR~\cite{fan2025incentivizing} proposed process-based rewards to incentivize consistent and scalable thinking chains. Our work builds on these by addressing semantic decoupling through a hybrid reasoning similarity reward and employing GDPO~\cite{liu2026gdpo} to prevent signal collapse during multi-reward optimization.


\begin{figure*}[t]
    \centering
    \includegraphics[width=0.95\textwidth]{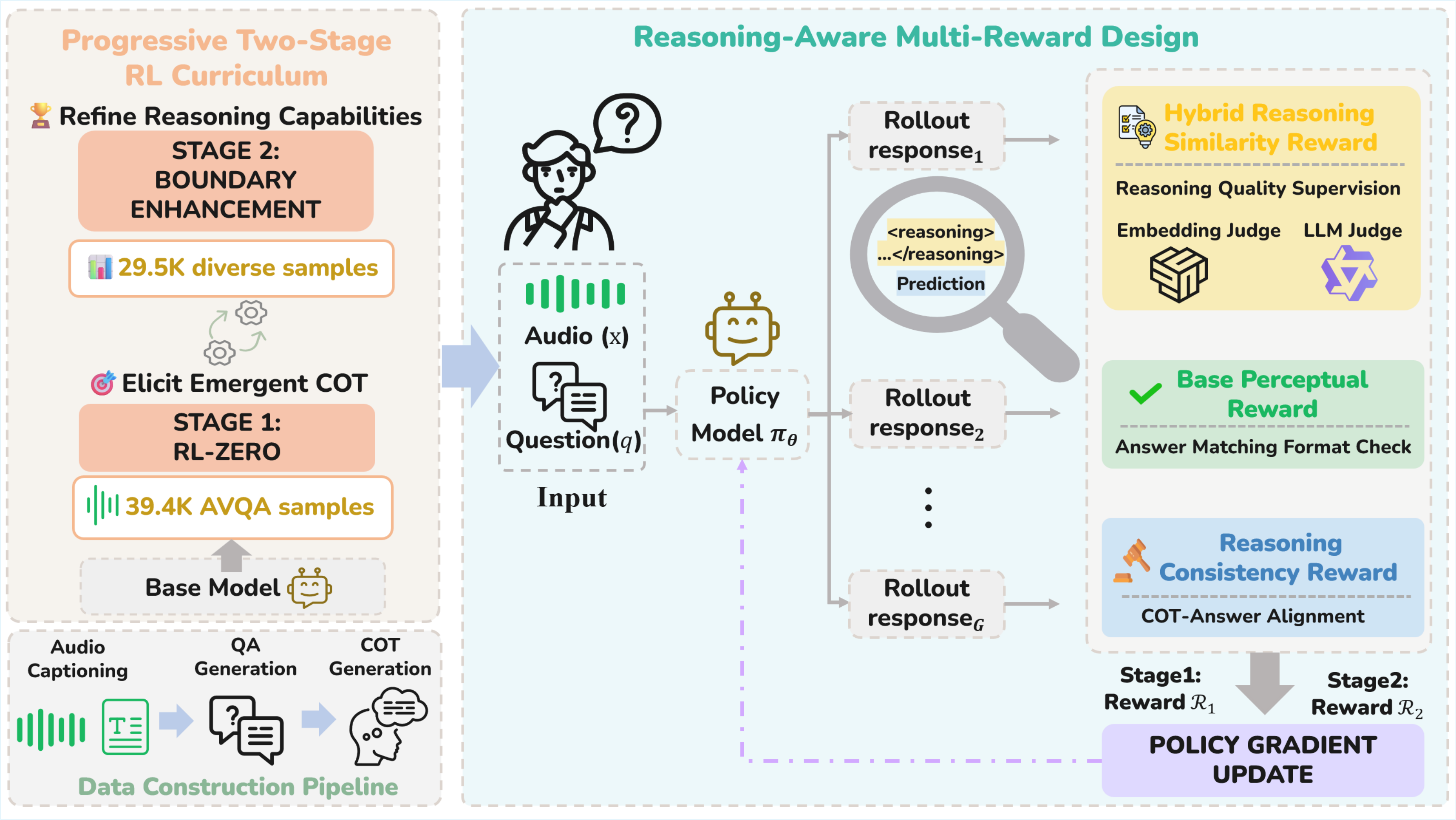}
    \caption{Overview of Audio-DeepThinker. \textbf{Bottom-left:} Data construction pipeline generating reference CoT annotations via audio captioning, QA generation, and CoT generation. \textbf{Top-left:} Progressive two-stage RL curriculum: Stage~1 elicits reasoning on foundational audio QA via pure RL exploration, and Stage~2 refines it on acoustically challenging boundary cases. \textbf{Right:} Reasoning-aware multi-reward design combining hybrid reasoning similarity reward, base perceptual reward, and reasoning consistency reward.}
    \label{fig:method}
\end{figure*}

\begin{figure}[t]
    \centering
    \includegraphics[width=0.9\columnwidth]{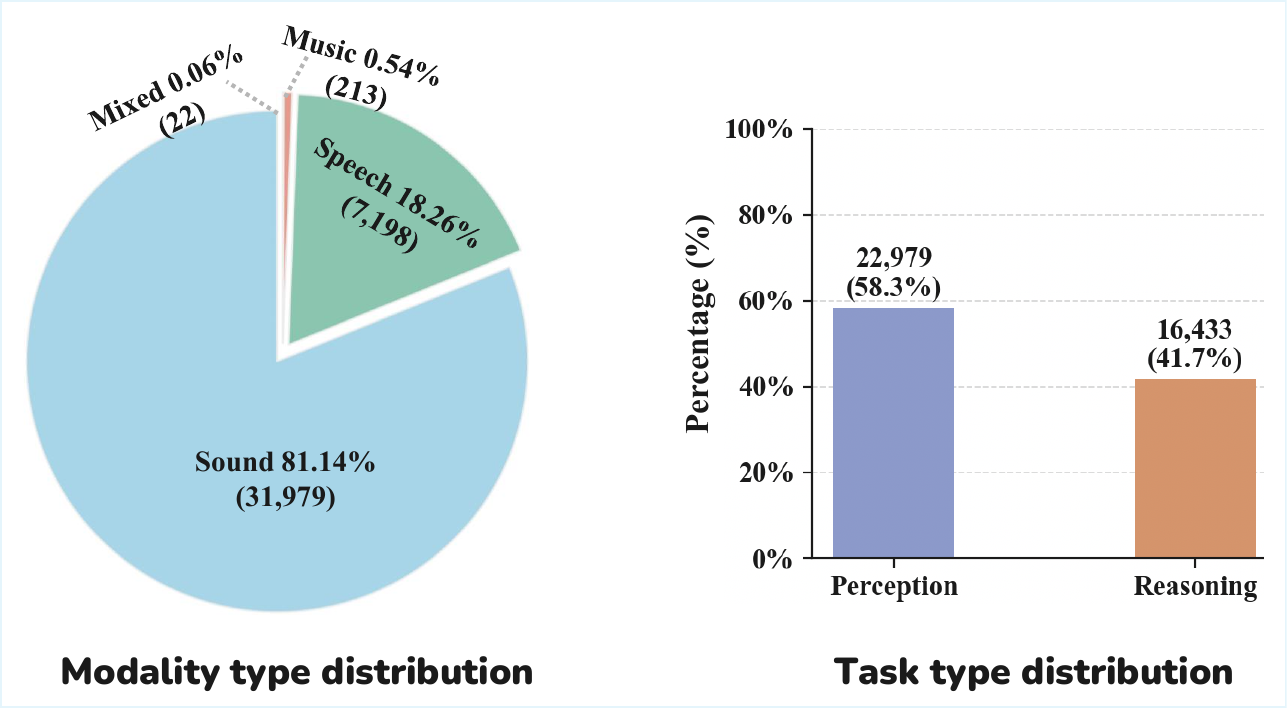}
    \vspace{0.5em}
    \includegraphics[width=0.9\columnwidth]{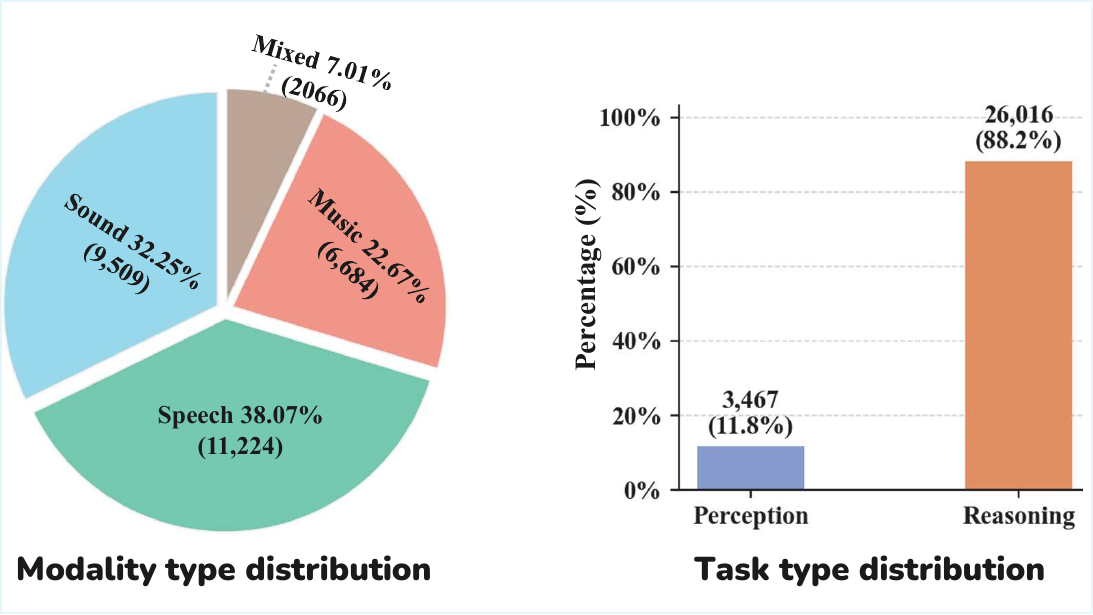}
    \caption{Training data distributions for the two RL stages. \textbf{Top:} Stage~1 dataset $\mathcal{D}_1$ (39,412 samples from AVQA), dominated by sound-event audio (81.14\%) with a roughly balanced split between perception (58.3\%) and reasoning (41.7\%) tasks. \textbf{Bottom:} Stage~2 dataset $\mathcal{D}_2$ (29,483 samples from diverse sources), featuring a substantially more balanced modality distribution across speech, sound, music, and mixed categories, with a deliberate emphasis on reasoning-intensive tasks (88.2\%).}
    \label{fig:data_distribution}
\end{figure}

\section{Method}
\label{sec:method}

We present Audio-DeepThinker, a framework that enables chain-of-thought reasoning capabilities to emerge in Large Audio-Language Models through reinforcement learning, without any supervised reasoning data. Our approach builds on a key insight: just as DeepSeek-R1~\cite{guo2025deepseek} demonstrated that reasoning can emerge naturally through RL exploration in text-only models, similar principles can be applied to the audio domain with carefully designed rewards and a progressive training curriculum.

\subsection{Overview}

As illustrated in Fig.~\ref{fig:method}, our framework consists of three components: (1) a data construction pipeline that generates high-quality CoT annotations without human labeling, (2) a reasoning-aware multi-reward design centered on a hybrid similarity reward providing fine-grained supervision on both answer correctness and reasoning quality, and (3) a progressive two-stage training paradigm that first establishes foundational reasoning patterns, then refines them on challenging cases.

\subsection{Data Construction Pipeline}
\label{ssec:data}

A practical challenge in training audio reasoning models is the scarcity of high-quality chain-of-thought annotations. Rather than relying on expensive human labeling, we construct training data through an automated three-step pipeline. Each training sample ultimately consists of a question $q$, a ground-truth answer $a^*$, and a reference reasoning chain $r^*$.

\textbf{Step 1: Audio Captioning.} Given an audio input $x$, we use Qwen3-Omni-Captioner~\cite{xu2025qwen3} $f_{\text{cap}}$ to generate a detailed textual description $c = f_{\text{cap}}(x)$. The caption $c$ captures key acoustic characteristics, including sound events, musical elements, and speech content.

\textbf{Step 2: Question-Answer Generation.} For datasets that lack question-answer annotations (such as AudioSet~\cite{gemmeke2017audio} and MagnaTagATune~\cite{law2009evaluation}), we employ Qwen3-235B-A22B-Instruct-2507~\cite{qwen3technicalreport} to generate $(q, a^*) = f_{\text{qa}}(c)$ from the caption; for datasets that already provide question-answer pairs (such as AVQA~\cite{10.1145/3503161.3548291}), we directly adopt the original $(q, a^*)$.

\textbf{Step 3: Chain-of-Thought Generation.} Based on the obtained question-answer pair $(q, a^*)$ and the caption $c$, we use DeepSeek V3.1~\cite{deepseekai2024deepseekv3technicalreport} to generate a reference reasoning chain: $r^* = f_{\text{cot}}(c, q, a^*)$. The detailed prompt template is provided in Appendix~\ref{ssec:cot_prompt}.

Through this pipeline, we obtain two stage-specific training sets. $\mathcal{D}_1$ consists of 39{,}412 samples from AVQA~\cite{10.1145/3503161.3548291}, providing well-structured queries for initial reasoning acquisition. $\mathcal{D}_2$ comprises 29{,}483 samples drawn from AudioMCQ~\cite{he2025measuring}, AudioSet~\cite{gemmeke2017audio}, MagnaTagATune~\cite{law2009evaluation}, Switchboard, MusicBench~\cite{melechovsky2024mustango}, CochlScene, MusicAVQA, and IEMOCAP~\cite{busso2008iemocap}, covering diverse audio tasks to strengthen reasoning generalization on challenging boundary cases. The detailed modality and task type distributions for both stages are visualized in Fig.~\ref{fig:data_distribution}.

\begin{figure*}[t]
    \centering
    \includegraphics[width=0.95\textwidth]{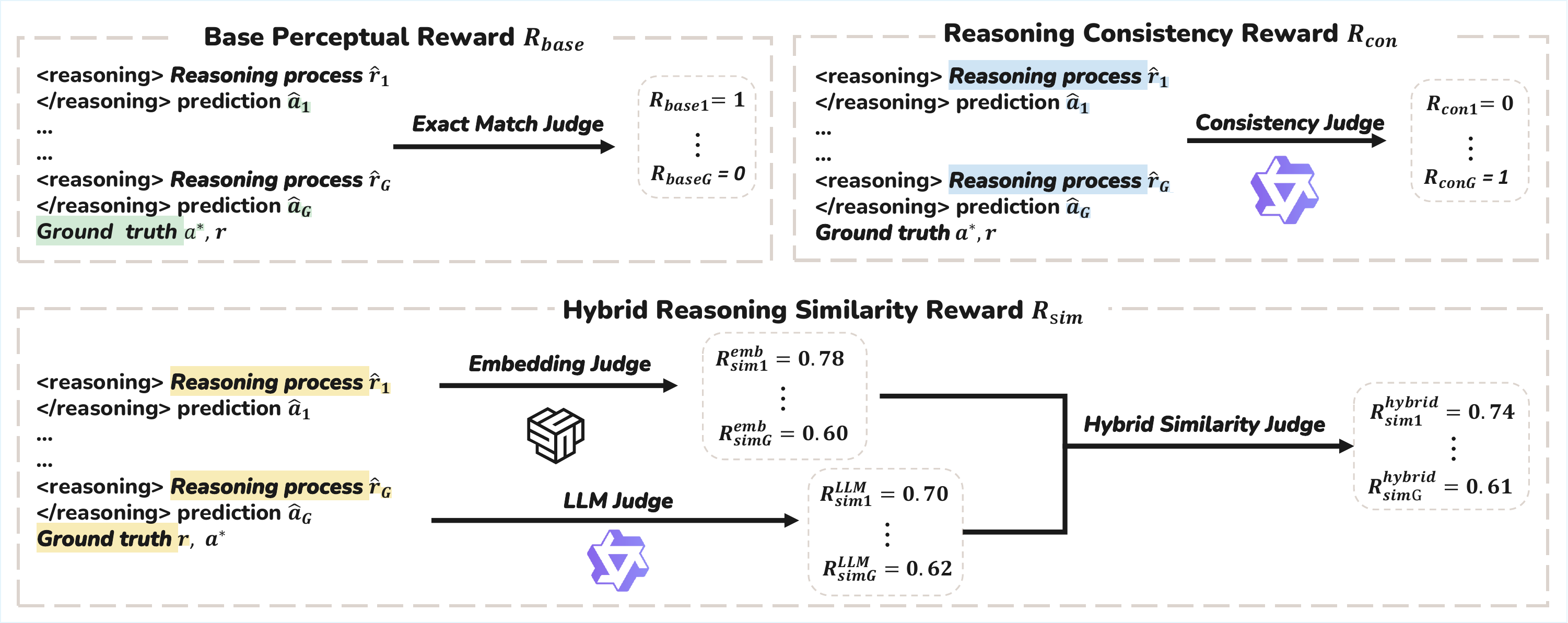}
    \caption{Detailed illustration of the three reward components. \textbf{Top-left:} Base perceptual reward $R_{\text{base}}$, combining answer correctness via exact matching and format compliance. \textbf{Top-right:} Reasoning consistency reward $R_{\text{con}}$, which uses an LLM judge to verify that the generated reasoning chain logically supports the predicted answer. \textbf{Bottom:} Hybrid reasoning similarity reward $R_{\text{sim}}^{\text{hybrid}}$, which evaluates the generated reasoning against the reference chain through an embedding judge (BGE-M3 cosine similarity) and an LLM judge (logical path evaluation), aggregating both scores into a final hybrid similarity score.}
    \label{fig:reward}
\end{figure*}

\subsection{Reasoning-Aware Multi-Reward Design}
\label{ssec:reward}

As illustrated in Fig.~\ref{fig:reward}, we propose a reasoning-aware multi-reward that jointly supervises answer correctness and the quality of the reasoning process. During RL training, given a question $q$, the policy model generates a reasoning chain $\hat{r}$ and an answer $\hat{a}$. The reward is computed by comparing $(\hat{r}, \hat{a})$ against the ground-truth answer $a^*$ and reference reasoning chain $r^*$.

\textbf{Base Perceptual Reward.} We define the base reward $R_{\text{base}}$ as the combination of answer correctness and output formatting:
\begin{equation}
    R_{\text{base}}(\hat{r}, \hat{a}) = \underbrace{R_{\text{acc}}(\hat{a}, a^*)}_{\text{correctness}} + \underbrace{R_{\text{fmt}}(\hat{r}, \hat{a})}_{\text{format}},
\end{equation}
where $R_{\text{acc}}(\hat{a}, a^*) = \mathbf{1}[\hat{a} = a^*]$ is a binary correctness indicator, and $R_{\text{fmt}}(\hat{r}, \hat{a}) = \mathbf{1}[(\hat{r}, \hat{a}) \in \mathcal{T}]$ verifies that the output conforms to the expected tag structure $\mathcal{T}$ (\textit{i.e.,} \texttt{<reasoning>}...\texttt{</reasoning>} followed by the answer).

\textbf{Reasoning Consistency Reward.} Following~\cite{wu2025audio}, we incorporate a consistency reward $R_{\text{con}}(\hat{r}, \hat{a}) = \psi(\hat{r}, \hat{a})$ in the first training stage to encourage stable reasoning patterns, where $\psi(\hat{r}, \hat{a}) \in \{0, 1\}$ is an LLM judge assessing whether the reasoning chain logically supports the predicted answer. This reward penalizes contradictions between the generated reasoning chain and the final answer, ensuring that the model's intermediate steps are coherent with its conclusion.

\textbf{Reasoning Similarity Reward.} To address the core limitation that existing methods lack fine-grained supervision over the content of generated reasoning chains, we propose a hybrid similarity reward that directly evaluates the quality of $\hat{r}$ against the reference $r^*$. It consists of two complementary components:

\textit{LLM-based Logical Evaluation.} We employ an LLM evaluator $\phi$ (Qwen3-235B-A22B-Instruct~\cite{qwen3technicalreport}) to assess reasoning quality along four dimensions: logical path alignment, key step coverage, reasoning strategy consistency, and analysis depth (prompt in Appendix~\ref{ssec:sim_prompt}):
\begin{equation}
    R_{\text{sim}}^{\text{LLM}}(\hat{r}, r^*) = \phi\bigl( \text{prompt}(\hat{r}, r^*) \bigr).
\end{equation}

\textit{Embedding-based Semantic Alignment.} We use BGE-M3~\cite{chen2024bge} to compute dense semantic similarity, providing a stable anchor signal during the critical reasoning emergence phase:
\begin{equation}
    R_{\text{sim}}^{\text{emb}}(\hat{r}, r^*) = \cos\bigl( \mathbf{e}(\hat{r}),\, \mathbf{e}(r^*) \bigr),
\end{equation}
where $\mathbf{e}(\cdot)$ denotes the BGE-M3 embedding.

In Stage~1, the hybrid similarity reward combines both components with equal weights:
\begin{equation}
    R_{\text{sim}}^{\text{hybrid}}(\hat{r}, r^*) = \frac{1}{2} R_{\text{sim}}^{\text{LLM}}(\hat{r}, r^*) + \frac{1}{2} R_{\text{sim}}^{\text{emb}}(\hat{r}, r^*),
\end{equation}
where both $R_{\text{sim}}^{\text{LLM}}$ and $R_{\text{sim}}^{\text{emb}}$ are bounded in $[0, 1]$.

In Stage~2, we switch to $R_{\text{sim}}^{\text{LLM}}$ only, removing the embedding anchor to grant the model greater flexibility to explore diverse reasoning strategies on challenging boundary cases. The rationale is that embedding cosine similarity penalizes reasoning chains that differ superficially from the reference, even when they follow an equally valid alternative logic. The LLM evaluator, by contrast, can recognize logically equivalent reasoning across different surface forms. In both stages, the similarity reward is applied only when the answer is correct ($\mathbf{1}[\hat{a} = a^*]$), ensuring that the model is incentivized to develop reasoning paths that are not only plausible but also lead to factually accurate conclusions.

\subsection{Progressive Two-Stage Training}
\label{ssec:training}

Our framework does not require supervised CoT fine-tuning; instead, high-quality reasoning emerges through pure RL exploration, provided that the training data and reward are appropriately staged.

\textbf{Stage 1: Foundational Reasoning Elicitation.} We begin by training the base model on the foundational dataset $\mathcal{D}_1$. The full reward $\mathcal{R}_1 = R_{\text{acc}} + R_{\text{fmt}} + R_{\text{con}} + R_{\text{sim}}^{\text{hybrid}}$ combines all rewards. For each sample $\tau = (x, q, a^*, r^*) \sim \mathcal{D}_1$, the optimization objective is:
\begin{equation}
    \theta_1 = \arg\max_{\theta} \mathbb{E}_{\pi_{\theta}} \left[
        \mathcal{R}_1(\hat{r}, \hat{a}; \tau) - \beta D_{\text{KL}}(\pi_{\theta} \parallel \pi_{\text{ref}})
    \right],
\end{equation}
where $\pi_{\theta}$ denotes the policy model parameterized by $\theta$, $\pi_{\text{ref}}$ is the initial pretrained model serving as the reference, $\beta$ is the KL penalty coefficient, and $D_{\text{KL}}$ denotes the KL divergence~\cite{cover1999elements}. Starting from a model with no reasoning capability, RL exploration gradually discovers that generating intermediate reasoning steps leads to higher rewards. This emergent behavior is the key finding: the model learns to ``think out loud'' not because it was explicitly taught to do so, but because doing so improves its reward signal. The combined supervision from format, consistency, and similarity rewards provides dense guidance during this critical emergence phase.

\textbf{Stage 2: Boundary Enhancement.} Once basic reasoning patterns are established, we train on acoustically challenging boundary cases from diverse audio sources $\mathcal{D}_2$ with a streamlined reward $\mathcal{R}_2 = R_{\text{acc}} + R_{\text{sim}}^{\text{LLM}}$, where $R_{\text{sim}}^{\text{LLM}}$ retains only the LLM evaluator component of the hybrid reward. For each sample $\tau \sim \mathcal{D}_2$:
\begin{equation}
\theta_{2} = \arg\max_{\theta} \mathbb{E}_{\pi_{\theta}} \left[
    \mathcal{R}_2(\hat{r}, \hat{a}; \tau) - \beta D_{\text{KL}}(\pi_{\theta} \parallel \pi_{\theta_1})
\right].
\label{eq:stage2_obj_fixed}
\end{equation}

Note that the reference policy changes from $\pi_{\text{ref}}$ to $\pi_{\theta_1}$, allowing fine-tuning from the Stage~1 checkpoint. This stage targets samples where the model's Stage~1 capabilities are insufficient. By removing format and consistency constraints that the model has already mastered, the reward signal concentrates entirely on answer correctness and reasoning depth, enabling more effective exploration on acoustically ambiguous cases. The effectiveness of this progressive approach can be understood from a curriculum learning perspective: training directly on difficult samples from the outset would lead to unstable exploration and poor convergence.

\textbf{Multi-Reward Optimization with GDPO.} To avoid the reward collapse problem in standard GRPO~\cite{shao2024deepseekmath}, where summing rewards before normalization causes different reward combinations to yield identical advantages, we adopt Group Reward-Decoupled Normalization Policy Optimization (GDPO)~\cite{liu2026gdpo}, which normalizes each reward independently before aggregation. This decoupled design ensures each reward dimension contributes meaningfully to the gradient signal.

\section{Experiments}
\label{sec:exp}

\subsection{Experimental Setup}
\label{ssec:setup}

\noindent\textbf{Model.} We adopt Qwen3-Omni-30B-A3B-Instruct~\cite{xu2025qwen3} as our base model, which employs a mixture-of-experts (MoE) architecture with 30B total parameters and 3B active parameters per token. The model contains 48 transformer layers with 128 experts per layer. Notably, the model lacks inherent CoT reasoning capability prior to our RL training, making it suitable for evaluating whether high-quality reasoning chains can be elicited through our progressive reward design.

\noindent\textbf{Training Details.} Training is conducted using the SWIFT~\cite{zhao2024swiftascalablelightweightinfrastructure} framework integrated with Megatron-LM~\cite{megatron-lm} on 64 GPUs with asynchronous rollout generation. The global batch size is 224 with a micro batch size of 4. We use a learning rate of 1e-6, a KL coefficient $\beta=0.001$, and sample $G=8$ rollout responses per GDPO optimization step. The maximum sequence length is 4096 tokens with a maximum completion length of 1024 tokens. For distributed training, we employ tensor parallelism (TP=4), expert parallelism (EP=4), and pipeline parallelism (PP=2). Full training hyperparameters are provided in Appendix~\ref{sec:hyperparameters}.

\noindent\textbf{Training Data.} Stage~1 uses 39,412 samples from AVQA~\cite{10.1145/3503161.3548291} for initial reasoning acquisition. Stage~2 uses 29,483 samples combining AudioMCQ (20,656) and diverse open-source datasets (8,827) including AudioSet, MagnaTagATune, Switchboard, MusicBench, CochlScene, MusicAVQA, and IEMOCAP, targeting acoustically challenging boundary cases.

\noindent\textbf{Evaluation Benchmarks.} We evaluate on three complementary benchmarks:
\begin{itemize}
    \item \textbf{MMAR}~\cite{ma2025mmarchallengingbenchmarkdeep}: A challenging benchmark for deep audio reasoning, encompassing single-modality tasks (sound, music, speech) and mixed-modality tasks requiring cross-modal integration. We report accuracy and the Rubrics score from the Interspeech 2026 Audio Reasoning Challenge~\cite{ma2026interspeech}, which evaluates the logic and completeness of the reasoning chain at the instance level.
    \item \textbf{MMAU}~\cite{mmau2024}: A massive multi-task audio understanding benchmark covering sound, music, and speech, with both test-mini and full test sets.
    \item \textbf{MMSU}: A multi-modal speech understanding benchmark focusing on linguistics (semantics, phonology) and paralinguistics, evaluating both perceptual accuracy and higher-order reasoning capabilities.
\end{itemize}

\subsection{Main Results on MMAR and MMAU}
\label{ssec:main_results}

\begin{table*}[t]
    \centering
    \caption{Performance comparison on MMAR and MMAU benchmarks. MMAR results are reported on the test set. MMAU results report the average across Sound, Music, and Speech. Best results in \textbf{bold}, second-best \underline{underlined}.}
    \label{tab:main_results}
\resizebox{1.\linewidth}{!}{
    \begin{tabular}{l c c c c c c c c | c c}
        \toprule
        \multirow{3}{*}{\textbf{Model}} & \multicolumn{8}{c|}{\textbf{MMAR}} & \multicolumn{2}{c}{\textbf{MMAU}} \\
        \cmidrule(lr){2-9} \cmidrule(lr){10-11}
        & \multicolumn{3}{c}{Single Modality} & \multicolumn{4}{c}{Mixed Modalities} & \multirow{2}{*}{Avg} & \multirow{2}{*}{Test-Mini} & \multirow{2}{*}{Test} \\
        \cmidrule(lr){2-4} \cmidrule(lr){5-8}
        & Sound & Music & Speech & Sound-Music & Sound-Speech & Music-Speech & S-M-S & & & \\
        \midrule
        \multicolumn{11}{c}{\textit{Closed-source Models}} \\
        GPT-4o mini Audio~\cite{hurst2024gpt} & 38.79 & 35.92 & 58.84 & 45.45 & 60.09 & 57.32 & 50.00 & 50.60 & 53.00 & 51.03 \\
        GPT-4o Audio~\cite{hurst2024gpt} & 53.94 & 50.97 & 70.41 & 63.64 & 72.48 & 62.20 & \textbf{75.00} & 63.50 & 62.50 & 60.82 \\
        Gemini 2.0 Flash~\cite{universalaiassistant} & 61.21 & 50.97 & 72.11 & \textbf{81.82} & 72.48 & 65.85 & \underline{70.83} & 65.60 & 70.50 & 67.03 \\
        Gemini 2.5 Flash~\cite{comanici2025gemini} & 58.18 & 47.09 & 76.53 & 63.64 & 77.52 & \textbf{76.83} & 54.17 & 67.00 & 71.80 & 67.39 \\
        \midrule
        \multicolumn{11}{c}{\textit{Open-source Models}} \\
        Qwen2-Audio-Instruct~\cite{chu2024qwen2} & 33.33 & 24.27 & 32.31 & 9.09 & 31.19 & 30.49 & 25.00 & 30.00 & 59.60 & 57.40 \\
        Audio Flamingo 2~\cite{ghosh2025audio} & 24.85 & 17.48 & 20.75 & 18.18 & 26.61 & 23.17 & 8.33 & 21.90 & 62.40 & 61.06 \\
        Audio Flamingo 3~\cite{goel2025audio} & - & - & - & - & - & - & - & 58.50 & 73.30 & 72.42 \\
        Kimi-Audio~\cite{ding2025kimi} & - & - & - & - & - & - & - & 57.60 & 68.20 & 64.40 \\
        Qwen-2.5-Omni~\cite{xu2025qwen25omnitechnicalreport} & 58.79 & 40.78 & 59.86 & 54.55 & 61.93 & 67.07 & 58.33 & 56.70 & 71.50 & 71.00 \\
        Covo-Audio~\cite{wang2026covo} & 57.58 & 41.26 & 62.50 & 36.36 & 59.17 & 58.54 & 62.50 & 55.30 & 75.30 & 71.89 \\
        Step-Audio 2~\cite{wu2025step} & - & - & - & - & - & - & - & - & 77.58 & 73.86 \\
        Qwen3-Omni-30B-A3B-Thinking~\cite{xu2025qwen3} & 61.21 & 47.09 & \underline{77.89} & 72.73 & 77.98 & 68.29 & \underline{70.83} & 67.80 & 75.80 & 74.99 \\
        Qwen3-Omni-30B-A3B-Instruct~\cite{xu2025qwen3} & 62.42 & \underline{57.28} & 77.21 & \textbf{81.82} & \underline{79.82} & 64.63 & \underline{70.83} & \underline{70.10} & 77.80 & 74.57 \\
        \midrule
        \multicolumn{11}{c}{\textit{Open-source Models Finetuned with RL}} \\
        Audio-Reasoner~\cite{xie2025audio} & 43.64 & 33.50 & 32.99 & 45.45 & 42.66 & 31.71 & 25.00 & 36.80 & 67.70 & 63.78 \\
        Omni-R1~\cite{rouditchenko2025omni} & 63.60 & 51.50 & 62.20 & 72.70 & 65.10 & 62.20 & 70.80 & 61.20 & 77.00 & 75.00 \\
        Ke-Omni-R~\cite{zhao2025keomnir} & 63.64 & 47.09 & 62.93 & 63.64 & 68.35 & 67.07 & 45.84 & 60.90 & 74.60 & 68.71 \\
        CESAR~\cite{fan2025incentivizing} & \underline{66.06} & 55.83 & 62.24 & 63.64 & 67.43 & 60.98 & 66.67 & 62.70 & 77.10 & - \\
        Audio-Thinker~\cite{wu2025audio} & \textbf{68.48} & 53.88 & 64.29 & \underline{72.73} & 71.56 & \underline{73.17} & 66.67 & 65.30 & 78.00 & 75.39 \\
        AudioMCQ~\cite{he2025measuring} & - & - & - & - & - & - & - & 67.10 & \underline{78.20} & \textbf{75.60} \\
        \rowcolor{highlight} \textbf{Audio-DeepThinker (Ours)} & \underline{66.06} & \textbf{64.08} & \textbf{80.27} & \textbf{81.82} & \textbf{80.73} & \underline{73.17} & \textbf{75.00} & \textbf{74.00} & \textbf{78.50} & \underline{75.44} \\
        \bottomrule
    \end{tabular}}
\end{table*}

Table~\ref{tab:main_results} presents the performance comparison on the MMAR and MMAU benchmarks. On MMAR, Audio-DeepThinker achieves 74.0\% average accuracy, establishing a new state-of-the-art among all open-source methods and surpassing several closed-source models. Compared to the base model Qwen3-Omni-Instruct (70.10\%), our method yields a +3.9\% absolute improvement, demonstrating that RL training can effectively enhance audio reasoning capabilities well beyond those acquired during pre-training. The most substantial gains appear in the \textbf{Music} category (+6.80\% over the base Instruct model), a domain that demands particularly nuanced reasoning about harmony, tempo, instrumentation, and genre attributes. This suggests that our progressive RL training is especially effective for tasks requiring multi-faceted acoustic analysis, where explicit CoT reasoning helps the model systematically decompose complex musical judgments. Our model also achieves consistent improvements across all \textbf{mixed-modality} tasks, with notable gains of +8.54\% on Music-Speech over the base Instruct model, indicating that the emergent CoT reasoning facilitates the integration of information across heterogeneous audio modalities, a capability that is difficult to acquire through pre-training alone.

On MMAU, our method achieves 78.50\% on test-mini, ranking first among all models including both open-source and closed-source systems, and 75.44\% on the full test set, which is highly competitive with the top-performing AudioMCQ (75.60\%). Notably, our model attains the \textbf{best Speech performance} on test-mini (80.78\%) and the second-best on the full test set (77.87\%), demonstrating particularly strong reasoning capabilities on speech understanding tasks that require comprehensive understanding of linguistic content, speaker characteristics, and contextual cues.

Across both benchmarks, Audio-DeepThinker substantially outperforms all existing RL-based methods (Audio-Thinker: 65.30\%/78.00\%, AudioMCQ: 67.10\%/78.20\%, CESAR: 62.70\%/77.10\% on MMAR/MMAU-test-mini), validating the joint effectiveness of the progressive training paradigm and the GDPO-based multi-reward optimization. This is noteworthy given that our method does not rely on any supervised CoT fine-tuning, further confirming the effectiveness of the SFT-free RL paradigm.

\subsection{Main Results on MMSU}
\label{ssec:mmsu_results}

\begin{table*}[t]
    \centering
    \caption{Performance comparison on the MMSU benchmark across perception and reasoning dimensions in Semantics (Seman.), Phonology (Phono.), and Paralinguistics (Para.) domains. Top two results are highlighted in \textbf{bold} and \underline{underline}, respectively.}
\resizebox{0.98\linewidth}{!}{
\begin{tabular}{l ccccccccc}
\toprule
\textbf{Models} & \multicolumn{4}{c}{\textbf{Perception (\%$\uparrow$)}} & \multicolumn{4}{c}{\textbf{Reasoning (\%$\uparrow$)}} & \textbf{Avg (\%$\uparrow$)} \\
\cmidrule(lr){2-5} \cmidrule(lr){6-9} \cmidrule(lr){10-10}
& \textbf{Seman.} & \textbf{Phono.} & \textbf{Para.} & \textbf{Avg} & \textbf{Seman.} & \textbf{Phono.} & \textbf{Para.} & \textbf{Avg} & \textbf{All} \\
\midrule
\multicolumn{10}{c}{\textit{Closed-source Models}} \\
GPT-4o-Audio & 59.70 & 41.56 & 21.44 & 39.67 & 80.83 & 78.74 & 26.25 & 71.96 & 56.38 \\
Gemini-1.5-Pro & 57.06 & 53.60 & 31.23 & 46.10 & 79.47 & 83.46 & 46.33 & 76.16 & 60.68 \\
Gemini-2.0-Flash & 47.17 & 41.30 & 30.62 & 40.83 & 70.69 & 70.69 & 36.16 & 47.83 & 51.03 \\
\midrule
\multicolumn{10}{c}{\textit{Open-source Models}} \\
SALMONN & 31.55 & 29.08 & 28.71 & 29.83 & 36.43 & 26.22 & 25.26 & 30.04 & 30.01 \\
GLM-4-Voice & 27.80 & 24.52 & 27.34 & 26.18 & 46.10 & 48.16 & 44.35 & 46.76 & 35.51 \\
Qwen2-Audio-Instruct & 52.14 & 32.87 & 35.56 & 39.02 & 77.62 & 64.81 & 46.67 & 68.90 & 53.27 \\
Kimi-Audio & 57.64 & 42.30 & 35.74 & 43.52 & 81.77 & 76.65 & \textbf{55.22} & 76.03 & 59.28 \\
Baichuan-Omni & 47.14 & 36.01 & 28.49 & 35.42 & 71.19 & 73.67 & 43.28 & 67.19 & 50.58 \\
Qwen2.5-Omni & 55.12 & 37.33 & 39.35 & 42.50 & 88.00 & 81.37 & 48.36 & 79.83 & 60.57 \\
Covo-Audio & 57.01 & 52.78 & 75.15 & 58.95 & 82.58 & 79.22 & 36.42 & 74.83 & 66.64 \\
Qwen3-Omni-Instruct & \underline{76.06} & \underline{56.68} & \textbf{80.99} & 70.97 & \underline{90.70} & 84.95 & \underline{52.84} & \underline{83.14} & 76.86  \\
Qwen3-Omni-Thinking & \textbf{80.47} & 55.61 & 80.30 & \underline{71.40} & 89.53 & \underline{85.67} & 51.64 & 82.73 & \underline{76.88}  \\
\midrule
\multicolumn{10}{c}{\textit{Open-source Models Finetuned with RL}} \\
Audio-Reasoner & - & - & - & - & - & - & - & - & 49.20  \\
R1-AQA & - & - & - & - & - & - & - & - & 61.60  \\
AudioMCQ & - & - & - & 63.10 & - & - & - & 78.90 & 70.70  \\
\rowcolor{highlight} \textbf{Audio-DeepThinker (Ours)} & 73.07 & \textbf{60.86} & \underline{80.59} & \textbf{71.59} & \textbf{90.88} & \textbf{87.51} & 45.97 & \textbf{83.31} & \textbf{77.26}  \\
\bottomrule
\end{tabular}}
\label{tab:mmsu}
\end{table*}

Table~\ref{tab:mmsu} presents results on the MMSU benchmark, which provides a fine-grained evaluation across both perception and reasoning dimensions in linguistic and paralinguistic domains. Our method achieves 77.26\% overall accuracy, surpassing both the Instruct (76.86\%) and Thinking (76.88\%) variants of the base model. The improvements are particularly pronounced in the \textbf{Phonology} dimension, where we observe +4.18\% in perception and +2.56\% in reasoning over Qwen3-Omni-Instruct. This is especially significant because phonological reasoning demands fine-grained acoustic analysis of phonemic structures, stress patterns, and prosodic features. The model must attend to subtle acoustic cues and chain multiple perceptual observations into logically coherent judgments, which is precisely the type of capability that emerges through our progressive RL training with semantically grounded rewards. Furthermore, the reasoning component (83.31\%) exceeds both base model variants, confirming that the CoT reasoning acquired through SFT-free RL exploration transfers effectively to linguistically grounded tasks that were not explicitly targeted during training. Compared to other RL-based methods, our approach achieves a substantial +6.56\% improvement over AudioMCQ (70.70\%), underscoring the advantage of reasoning-aware reward design over methods that only optimize final answer quality.

\subsection{Ablation Study}
\label{ssec:ablation}

We conduct ablation studies to validate the two core design choices: the hybrid reasoning similarity reward (evaluated on MMAR), and the progressive two-stage training paradigm (evaluated on both MMAR and MMAU-test-mini). In addition to accuracy, we report the Rubrics score from the Interspeech 2026 Audio Reasoning Challenge~\cite{ma2026interspeech}, an instance-level metric that evaluates the logic and completeness of the reasoning chain.

\begin{table}[t]
    \centering
    \caption{Ablation study on reward components (Stage~1 only). ``--'' indicates the model does not produce reasoning chains.}
    \label{tab:ablation_reward}
\resizebox{0.99\columnwidth}{!}{
    \begin{tabular}{l|cc}
        \toprule
        \textbf{Configuration} & \textbf{Acc (\%)} & \textbf{Rubrics (\%)} \\
        \midrule
        Qwen3-Omni-Instruct (Baseline) & 70.10 & -- \\
        \midrule
        Base Reward ($R_{\text{acc}} + R_{\text{fmt}}$) & 70.50 & 49.17 \\
        + Audio-Thinker~\cite{wu2025audio} ($+ R_{\text{con}} + R_{\text{think}}$) & 71.70 & 57.44 \\
        + Audio-DeepThinker ($+ R_{\text{con}} + R_{\text{sim}}^{\text{hybrid}}$) & \textbf{73.10} & \textbf{64.33} \\
        \bottomrule
    \end{tabular}}
\end{table}

\noindent\textbf{Effect of Hybrid Reasoning Similarity Reward.}
Table~\ref{tab:ablation_reward} isolates the contribution of each reward component during Stage~1. Training with only the base reward ($R_{\text{acc}} + R_{\text{fmt}}$) yields a marginal +0.4\% accuracy gain over the vanilla baseline with limited reasoning quality (Rubrics: 49.17\%), confirming that sparse correctness signals alone are insufficient to elicit meaningful reasoning. Adding Audio-Thinker's rewards ($R_{\text{con}} + R_{\text{think}}$)~\cite{wu2025audio} improves accuracy to 71.70\%, but Rubrics remains limited at 57.44\%, indicating that while the model learns to generate reasoning chains, the quality remains suboptimal. This is precisely the semantic decoupling problem identified in Section~\ref{sec:intro}: the model produces structured outputs whose intermediate reasoning chains lack genuine grounding in the audio evidence. Replacing $R_{\text{think}}$ with our hybrid similarity reward $R_{\text{sim}}^{\text{hybrid}}$ yields substantial gains in both accuracy (+1.4\%, 71.70\%$\rightarrow$73.10\%) and reasoning quality (+6.89\%, 57.44\%$\rightarrow$64.33\%), demonstrating that directly supervising reasoning content through the combination of LLM-based logical evaluation and embedding-based semantic anchoring is far more effective than merely encouraging longer or more structured thinking.

\begin{table}[t]
    \centering
    \caption{Ablation on progressive two-stage training evaluated on MMAR and MMAU-test-mini, with generalization to Qwen3-Omni-Thinking. ``--'' indicates the metric is not applicable.}
    \label{tab:ablation_stage}
\resizebox{0.98\columnwidth}{!}{
    \begin{tabular}{l|cc|cc}
        \toprule
        & \multicolumn{2}{c|}{\textbf{Qwen3-Omni-Instruct}} & \multicolumn{2}{c}{\textbf{Qwen3-Omni-Thinking}} \\
        \cmidrule(lr){2-3} \cmidrule(lr){4-5}
        \textbf{Training} & \textbf{Acc (\%)} & \textbf{Rubrics (\%)} & \textbf{Acc (\%)} & \textbf{Rubrics (\%)} \\
        \midrule
        \multicolumn{5}{c}{\textit{MMAR}} \\
        Baseline (No RL) & 70.10 & -- & 68.10 & 56.95 \\
        Stage~1 only & 73.10 & 64.33 & 69.50 & 59.79 \\
        Stage~2 only & 73.40 & 62.53 & 68.70 & 61.55 \\
        Stage~1 $\rightarrow$ Stage~2 & \textbf{74.00} & \textbf{65.29} & \textbf{70.00} & \textbf{62.12} \\
        \midrule
        \multicolumn{5}{c}{\textit{MMAU-test-mini}} \\
        Baseline (No RL) & 77.80 & -- & 75.80 & -- \\
        Stage~1 only & 77.10 & -- & 76.40 & -- \\
        Stage~2 only & 77.30 & -- & 77.60 & -- \\
        Stage~1 $\rightarrow$ Stage~2 & \textbf{78.50} & -- & \textbf{78.00} & -- \\
        \bottomrule
    \end{tabular}}
\end{table}

\noindent\textbf{Effect of Progressive Two-Stage Training.}
Table~\ref{tab:ablation_stage} validates the progressive curriculum across two base models and two benchmarks. For both models, the full progressive pipeline (Stage~1 $\rightarrow$ Stage~2) consistently achieves the best results on all metrics, confirming that the two stages play complementary roles.

On MMAR, Stage~1 alone achieves strong reasoning quality for Qwen3-Omni-Instruct (Rubrics: 64.33\%), confirming that SFT-free RL successfully elicits foundational CoT without supervised reasoning data. Stage~2 alone yields comparable accuracy (73.40\%) but lower Rubrics (62.53\%), suggesting that training directly on boundary cases without foundational reasoning produces superficially correct but poorly grounded chains. The MMAU-test-mini results reveal a more pronounced pattern: for Qwen3-Omni-Instruct, neither Stage~1 alone (77.10\%) nor Stage~2 alone (77.30\%) surpasses the baseline (77.80\%), yet the progressive pipeline achieves 78.50\%. This indicates that individual stages may overspecialize on their respective training distributions, and only the progressive combination yields robust generalization across diverse audio understanding tasks. The same complementarity holds for Qwen3-Omni-Thinking, where the progressive pipeline consistently outperforms both individual stages on both benchmarks. Across all configurations, these results validate that the progressive curriculum is essential for reliable performance gains.

\noindent\textbf{Generalization to Other Base Models.}
Table~\ref{tab:ablation_stage} also demonstrates generalization of our framework to Qwen3-Omni-Thinking, a variant that already possesses built-in thinking capability. Even on this stronger baseline, our full progressive pipeline yields +1.9\% on MMAR (68.10\%$\rightarrow$70.00\%), +5.17 Rubrics improvement (56.95$\rightarrow$62.12), and +2.2\% on MMAU-test-mini (75.80\%$\rightarrow$78.00\%), demonstrating that the framework is architecture-agnostic and can serve as a general recipe for audio reasoning enhancement across different model configurations.

\section{Analysis}
\label{sec:analysis}

To gain deeper insights into how CoT reasoning capabilities are acquired through RL training, we conduct comprehensive interpretability analyses. We systematically compare the internal representations across the progressive training pipeline (Base $\rightarrow$ Stage~1 $\rightarrow$ Stage~2) at multiple levels of granularity: representation drift across layers, MoE expert dynamics and routing reorganization, and token-level prediction dynamics at reasoning boundaries.

\subsection{Representation Drift Analysis}

\begin{figure}[t]
    \centering
    \subfigure[Stage~1 vs.\ Base model]{%
        \includegraphics[width=0.48\columnwidth]{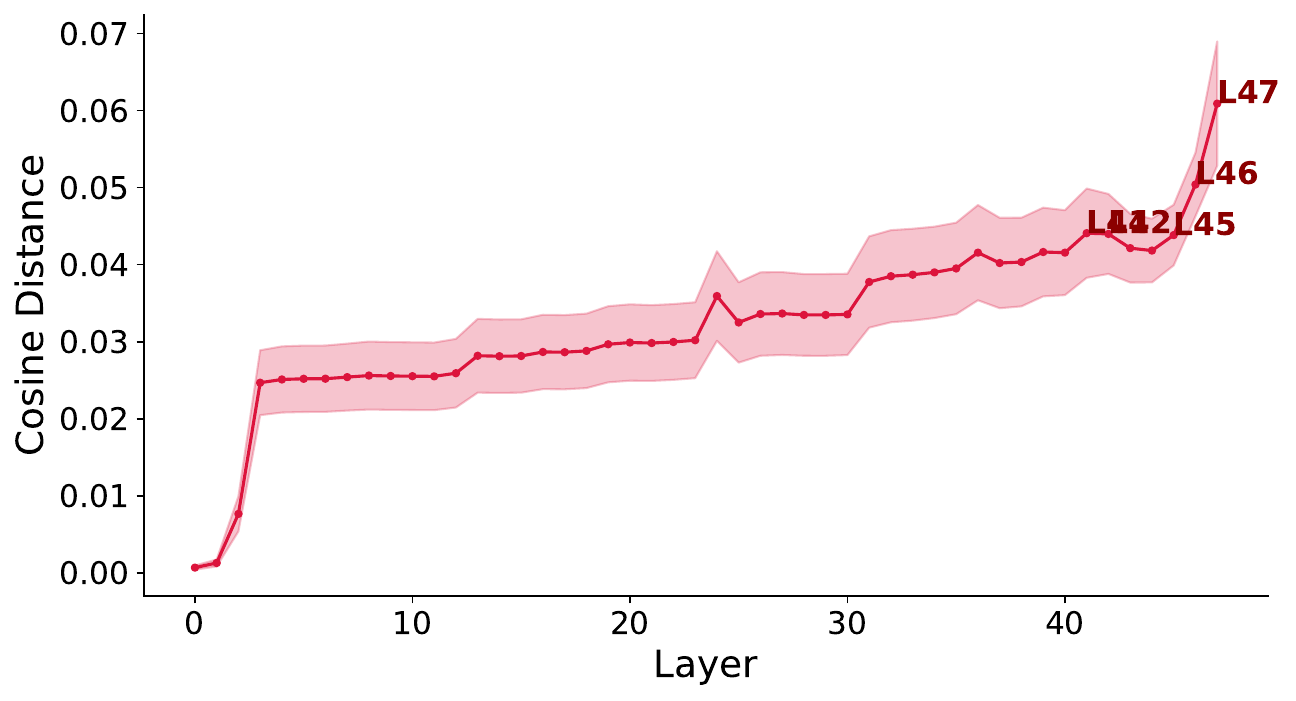}%
    }\hfill%
    \subfigure[Stage~2 vs.\ Stage~1]{%
        \includegraphics[width=0.48\columnwidth]{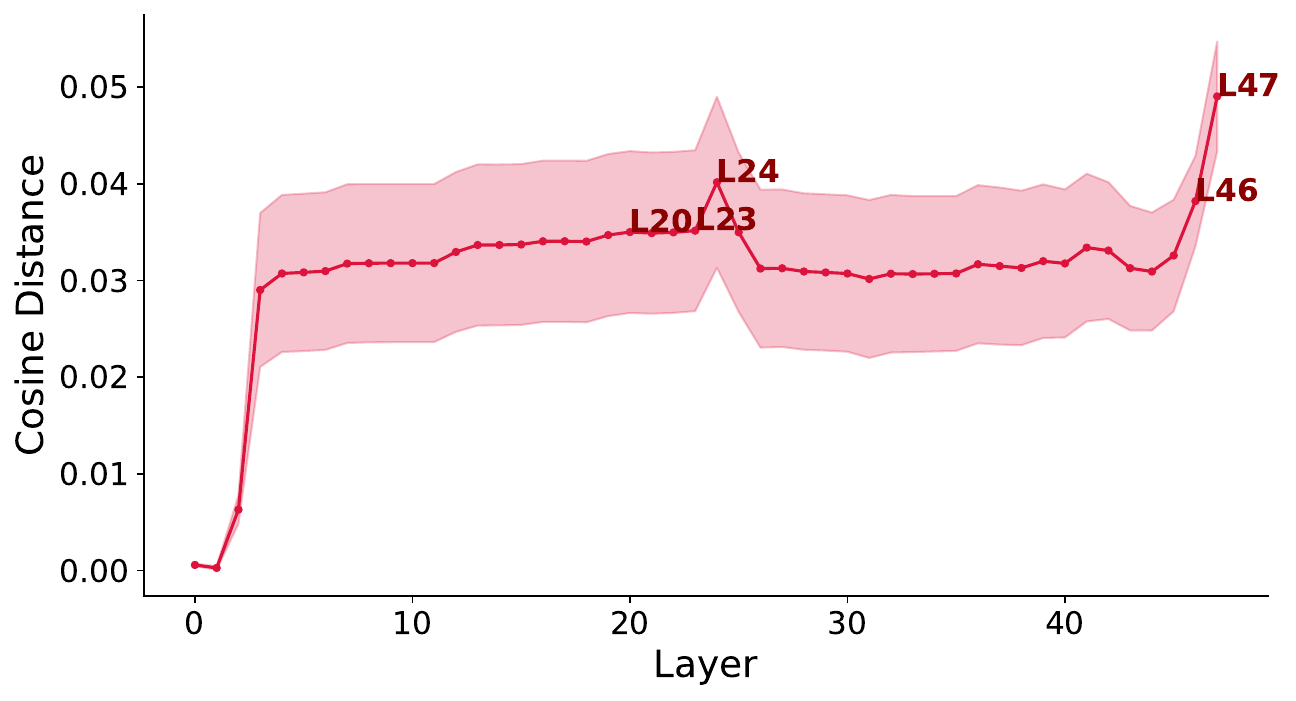}%
    }
    \caption{Progressive representation drift (cosine distance) across the two RL stages. \textbf{(a)} Stage~1 vs.\ Base model, showing a monotonically increasing drift that accelerates in L40+ (peaking at $\approx$0.061 at L47). \textbf{(b)} Stage~2 vs.\ Stage~1, revealing a qualitatively different bimodal pattern: a mid-layer plateau around L20--L24 ($\approx$0.035--0.040) in addition to the upper-layer spike at L47 ($\approx$0.049).}
    \label{fig:drift_analysis}
\end{figure}

Fig.~\ref{fig:drift_analysis} visualizes the progressive representation drift across the two RL training stages. For the same input, we extract hidden states from the compared model pair at each layer and compute cosine distance ($1 - \cos(\mathbf{h}_A, \mathbf{h}_B)$) to measure directional changes in the representation space.

\textbf{Stage~1 (Base $\rightarrow$ Stage~1)} creates a monotonically increasing drift that accelerates in the upper layers. Cosine distance rises gradually from $\approx$0.025 at L4 through $\approx$0.040 at L40, then spikes to $\approx$0.061 at L47. This pattern indicates that Stage~1's representational restructuring is distributed across all layers but concentrates its largest changes in the final computation stages where CoT reasoning patterns must be established.

\textbf{Stage~2 (Stage~1 $\rightarrow$ Stage~2)} exhibits a qualitatively different, bimodal drift pattern. In addition to the expected upper-layer spike at L47 ($\approx$0.049), Stage~2 introduces a distinct mid-layer plateau around L20--L24 (cosine $\approx$0.035--0.040), which then recedes to $\approx$0.030 through L28--L40 before the final spike. This mid-layer modification is absent in Stage~1, suggesting that refining performance on acoustically challenging boundary cases requires adjustments to intermediate representational processing, not just upper-layer generation patterns. The two stages thus reveal complementary modification strategies: Stage~1 primarily reshapes the upper-layer computation responsible for output decisions, while Stage~2 additionally adapts mid-layer representations, likely reflecting the need for finer-grained acoustic feature processing on boundary cases.

\subsection{MoE Expert Analysis}

\begin{figure*}[t]
    \centering
    \begin{minipage}{0.48\textwidth}
        \centering
        \includegraphics[width=\linewidth]{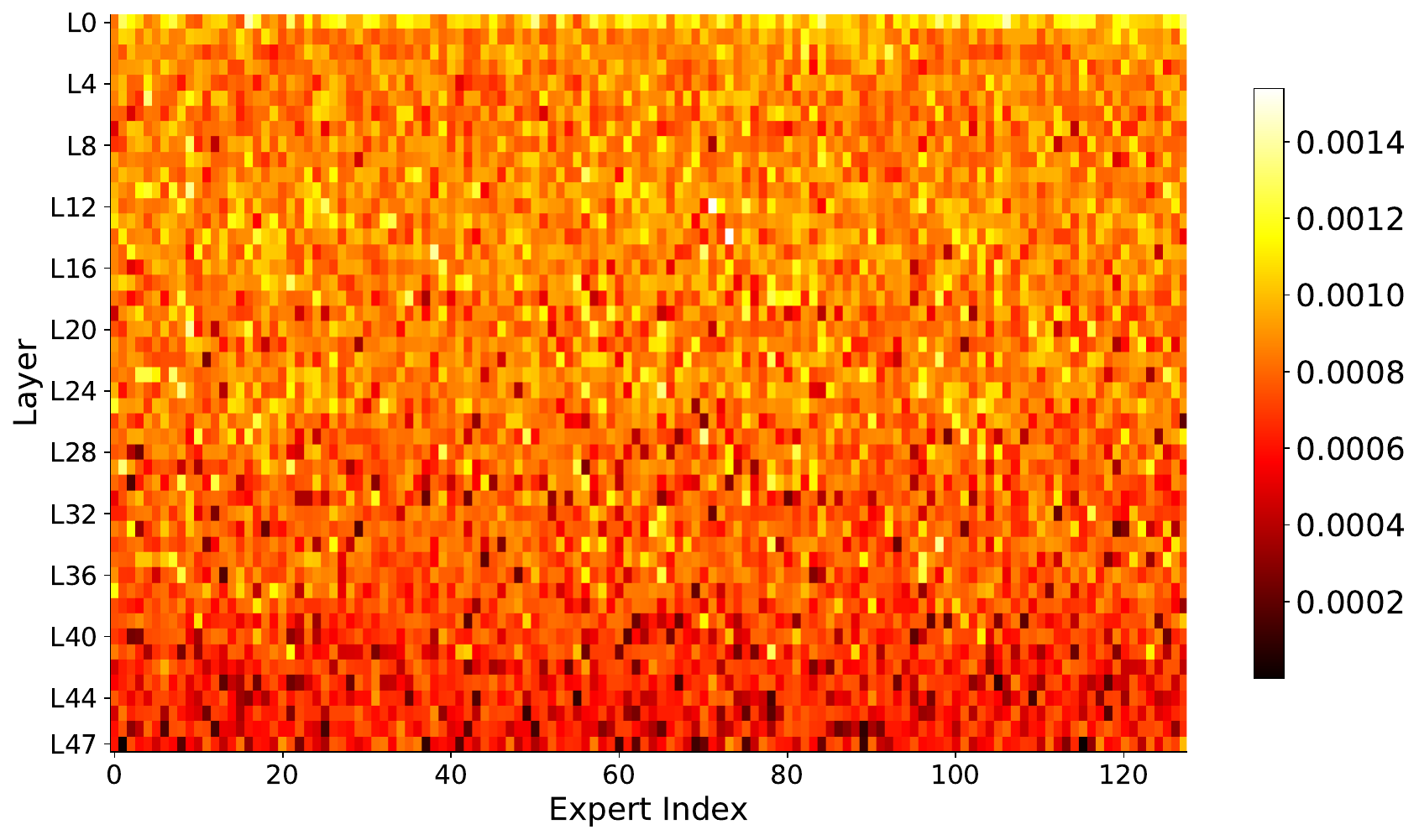}
    \end{minipage} \hfill
    \begin{minipage}{0.48\textwidth}
        \centering
        \includegraphics[width=\linewidth]{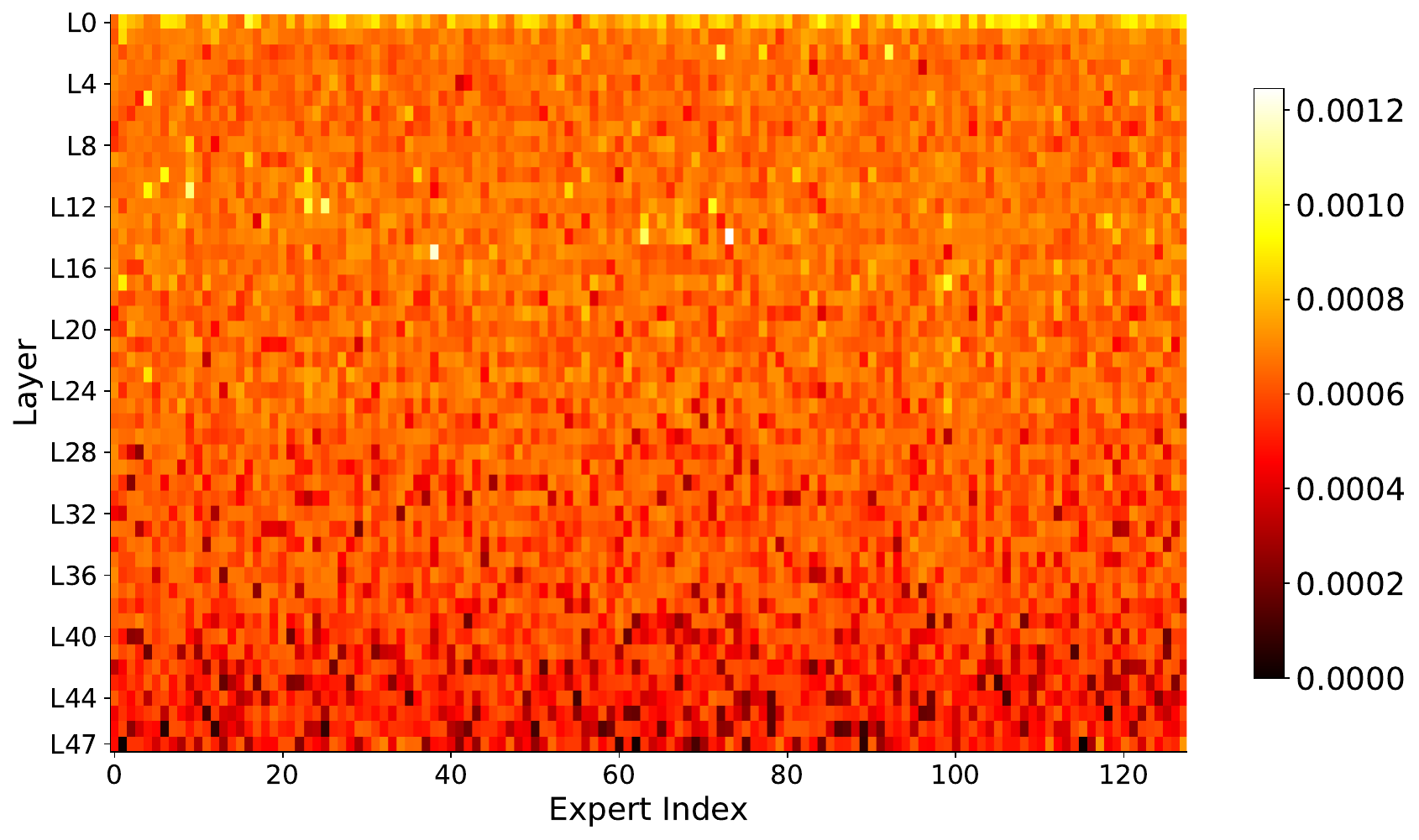}
    \end{minipage}
    \vspace{0.3em}
    \begin{minipage}{0.48\textwidth}
        \centering
        \includegraphics[width=\linewidth]{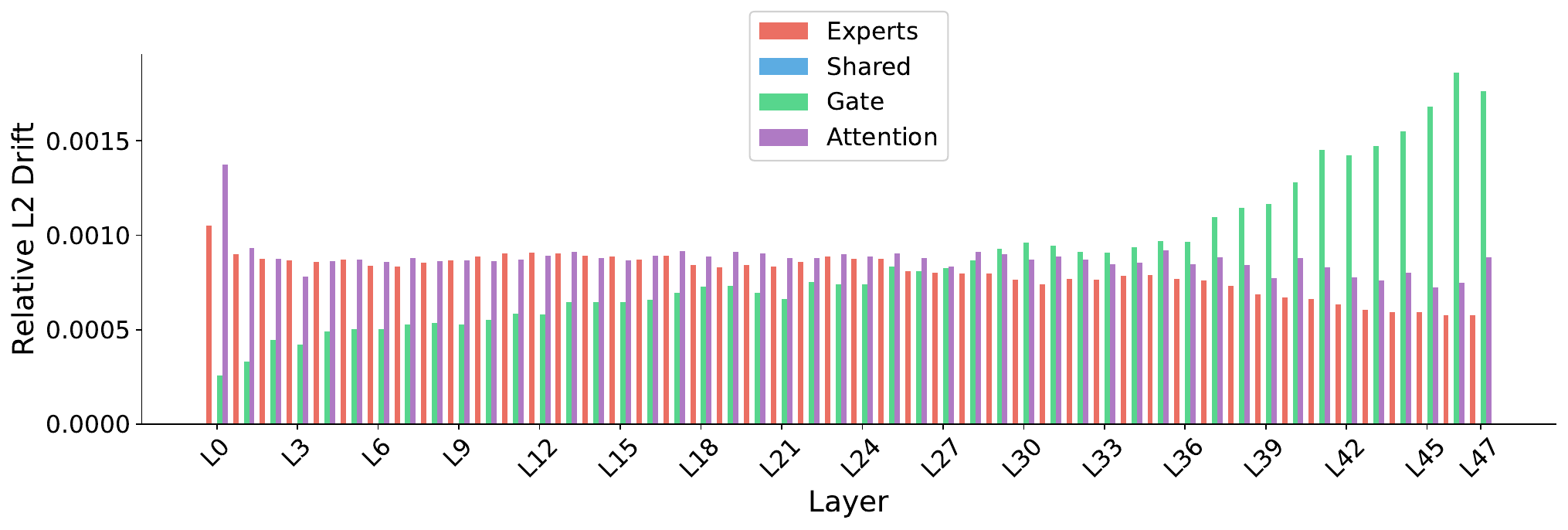}
    \end{minipage} \hfill
    \begin{minipage}{0.48\textwidth}
        \centering
        \includegraphics[width=\linewidth]{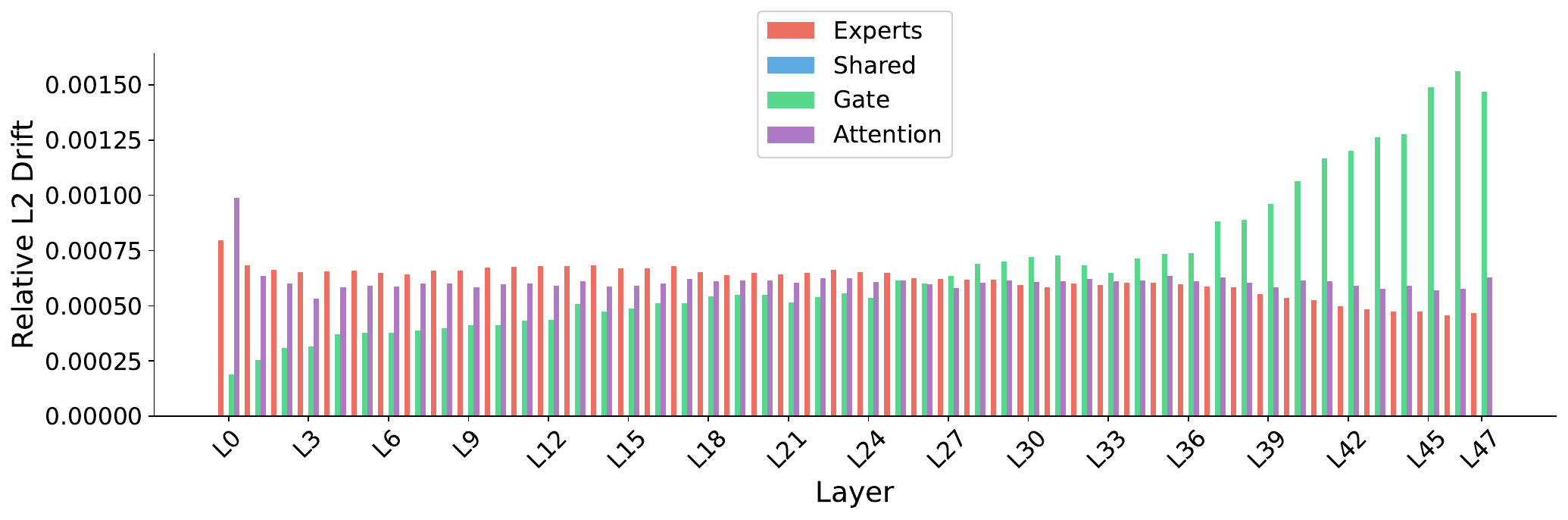}
    \end{minipage}
    \caption{Progressive MoE analysis across the two RL stages. \textbf{Top row:} Per-expert drift heatmaps (128 experts $\times$ 48 layers). \textbf{Bottom row:} Component-level drift decomposed into Experts, Shared, Gate, and Attention. \textbf{Left column:} Stage~1 vs.\ Base. \textbf{Right column:} Stage~2 vs.\ Stage~1. All drifts are measured as relative L2 norm $\lVert \mathbf{W}_{\mathrm{after}} - \mathbf{W}_{\mathrm{before}} \rVert_2 / \lVert \mathbf{W}_{\mathrm{before}} \rVert_2$ on raw weight matrices before any activation function. Both stages exhibit the same mechanistic signature: gating drift increases sharply in the upper layers while expert drift remains flat or decreases, indicating that RL progressively reshapes token routing rather than expert knowledge.}
    \label{fig:moe_analysis}
\end{figure*}

Since the base model employs a Mixture-of-Experts (MoE) architecture with 128 experts per layer, the MoE dynamics provide a unique lens into how RL training progressively reorganizes the model's computational routing to support reasoning. Fig.~\ref{fig:moe_analysis} presents per-expert drift heatmaps and component-level drift decompositions for both training stages.

All drifts are measured as the relative L2 norm of the parameter difference: $\lVert \mathbf{W}_{\mathrm{after}} - \mathbf{W}_{\mathrm{before}} \rVert_2 / \lVert \mathbf{W}_{\mathrm{before}} \rVert_2$, computed on the raw weight matrices before any activation function (e.g., softmax in gating), ensuring fair comparison across components with different parameter scales. This formulation avoids the confound where softmax output magnitudes could artificially scale gradient-based drift estimates.

\textbf{Stage~1 (Base $\rightarrow$ Stage~1).} The per-expert heatmap (top-left) shows that early layers (L0 to L16) exhibit relatively uniform and moderate drift (warm tones) across all experts, while deeper layers (L36 to L47) display increasingly heterogeneous patterns with large regions of near-zero drift (dark patches) interspersed with a sparse set of moderately drifted experts. This sparsity indicates that Stage~1 selectively reconfigures a task-relevant subset of experts for CoT generation.

The component-level drift (bottom-left) reveals the core mechanistic signature: gating network drift increases from $\sim$0.0003 at L0 to $\sim$0.0018 at L45, while expert drift decreases from $\sim$0.0011 at L0 to $\sim$0.0006 at L45+. At L45, gating drift reaches $\sim$3$\times$ higher than expert drift, revealing that Stage~1 primarily learns new token routing strategies rather than rewriting expert knowledge. The elevated attention drift at L0 ($\sim$0.0014) may reflect adjustments to initial information aggregation patterns.

\textbf{Stage~2 (Stage~1 $\rightarrow$ Stage~2).} The heatmap (top-right) shows a structurally consistent pattern: deep-layer sparsity persists with attenuated overall magnitudes (colorbar max 0.0012 vs.\ 0.0014). The component-level drift (bottom-right) confirms the same mechanistic signature: gating drift increases from $\sim$0.0002 at L0 to $\sim$0.0016 at L47 ($\sim$8$\times$), while expert drift declines from $\sim$0.0008 at L0 to $\sim$0.0005 at L47. This progressive consistency across both stages provides strong evidence that RL acquires reasoning capability through routing reorganization rather than knowledge rewriting, and that this mechanism operates incrementally across the progressive curriculum.

\subsection{Token Prediction Dynamics at Reasoning Boundaries}

\begin{figure}[ht]
    \centering
    \subfigure[Entropy comparison]{%
        \includegraphics[width=0.48\columnwidth]{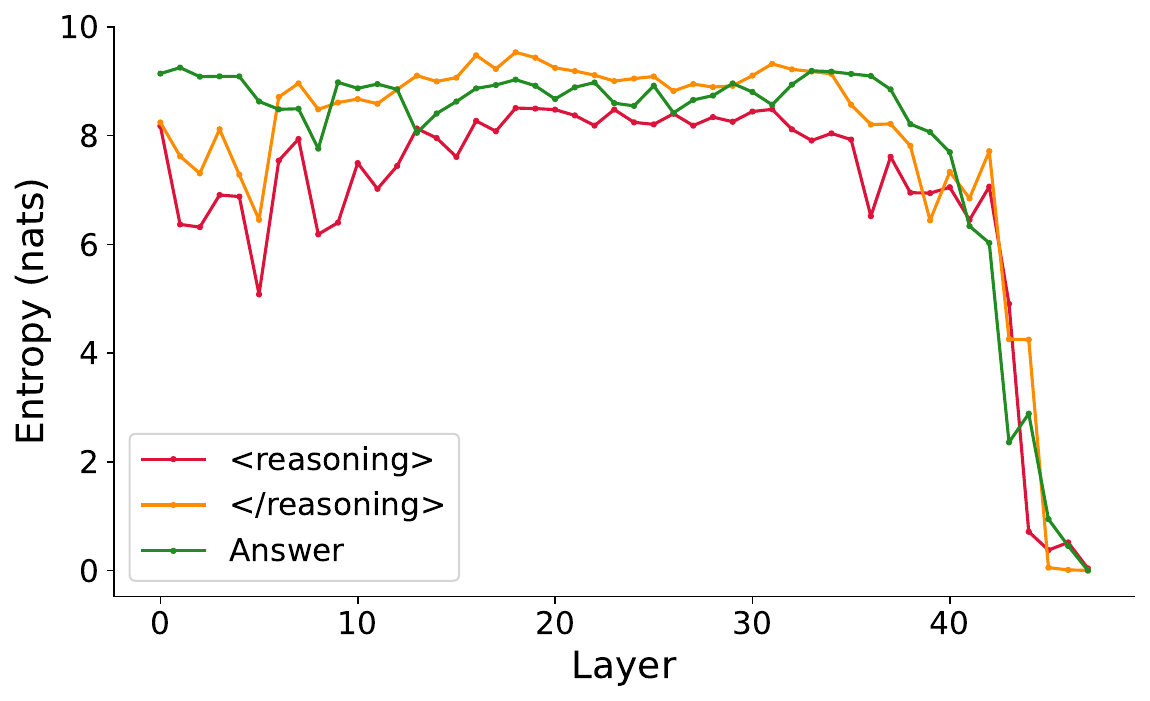}%
    }\hfill%
    \subfigure[Decision layer distribution]{%
        \includegraphics[width=0.48\columnwidth]{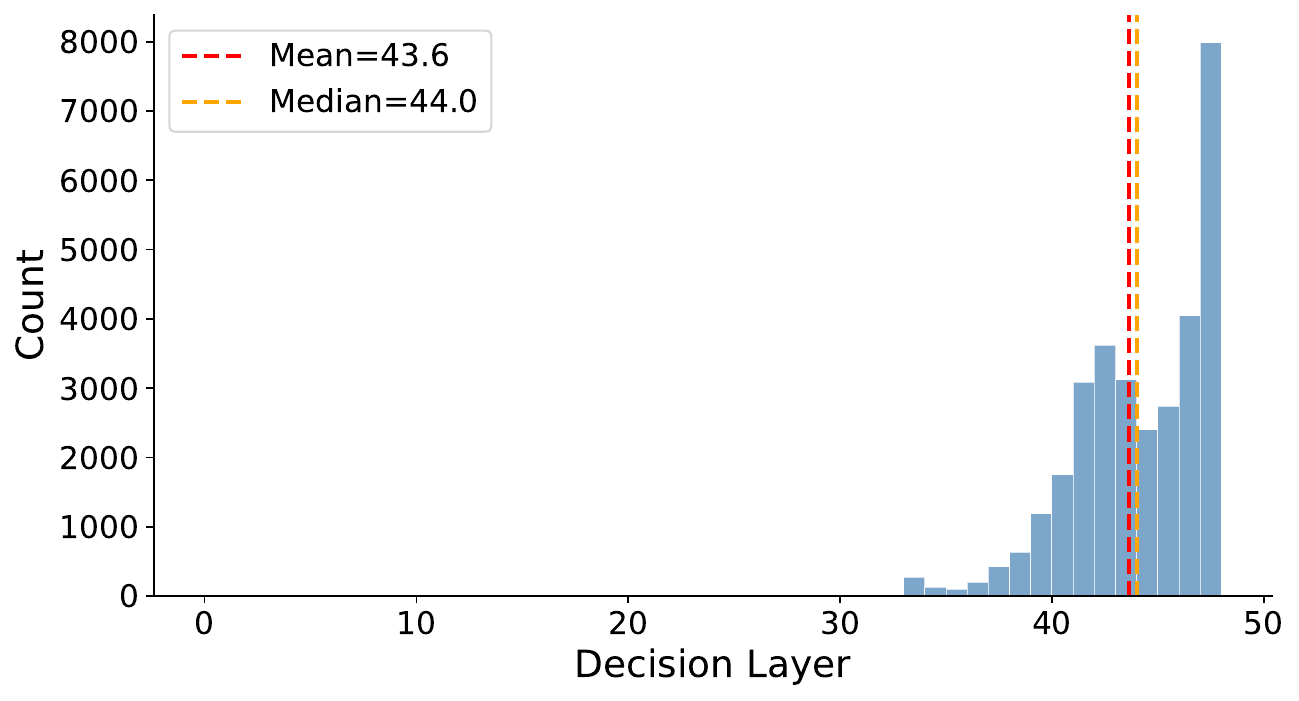}%
    }
    \caption{Token prediction dynamics via logit lens analysis. \textbf{(a)} Entropy comparison at three critical generation moments: the \texttt{<reasoning>} tag, the \texttt{</reasoning>} tag, and the final answer token. The \texttt{<reasoning>} token maintains the lowest entropy throughout, with all three tokens converging to near-zero in L42--L47. \textbf{(b)} Decision layer distribution across 34{,}998 generated tokens, confirming that output decisions concentrate in the upper layers.}
    \label{fig:entropy}
\end{figure}

To understand how the RL-trained model internally resolves reasoning-related generation decisions, we apply the logit lens technique~\cite{geva2022transformer}, projecting each layer's hidden state through the final layer norm and LM head to obtain next-token distributions. We analyze three critical generation moments: the opening \texttt{<reasoning>} tag, the closing \texttt{</reasoning>} tag, and the final answer token. Fig.~\ref{fig:entropy} presents two complementary perspectives.

\noindent\textbf{Entropy Dynamics.} The entropy comparison (Fig.~\ref{fig:entropy}a) reveals how the relative predictive uncertainty across the three token types evolves through the layers. The \texttt{<reasoning>} token consistently maintains the lowest entropy throughout the network, indicating that the model's distributional certainty about reasoning initiation is higher than for other token types even in early layers. All three tokens undergo a dramatic entropy collapse in \textbf{L42 to L47}, but \texttt{<reasoning>} drops earliest and most steeply, reaching near-zero by L45, while \texttt{</reasoning>} and Answer converge to near-zero at L47. This ordering reveals that the model resolves when to start reasoning before resolving the specific answer content, consistent with a hierarchical generation process where reasoning structure is decided before reasoning substance. Notably, the two-layer gap between the convergence of \texttt{<reasoning>} (L45) and the other two token types (L47) suggests that the model dedicates approximately two additional layers of computation to determine answer content after the reasoning mode has already been committed.

\noindent\textbf{Decision Layer Distribution.} The histogram of decision layers (Fig.~\ref{fig:entropy}b) across 34{,}998 generated tokens (31{,}682 with valid decision layers) exhibits a strongly right-skewed distribution concentrated in L35 to L47, with a mean of 43.6 and a median of 44.0. The pronounced peak at L47 indicates that a substantial fraction of tokens are resolved only at the final layer. Together with the representation drift analysis (Fig.~\ref{fig:drift_analysis}), these results converge on a consistent conclusion: RL training concentrates its modifications in the upper layers (L40+), which are precisely the layers where the model commits to generation decisions. This suggests that the reasoning capabilities acquired through RL are localized in a narrow computational band of upper transformer layers rather than being distributed throughout the network. The alignment between the decision layer concentration (L40+) and the gating drift peak (L45+, Fig.~\ref{fig:moe_analysis}) further indicates that RL training selectively modifies the layers where token-level output decisions are actively formed, leaving lower-layer feature extraction largely intact.

\subsection{Mechanistic Implications}

The above analyses converge on two findings with broader implications for RL fine-tuning of MoE-based language models. First, RL training reshapes routing strategies rather than expert knowledge: gating drift dominates expert drift in the upper layers across both stages (Fig.~\ref{fig:moe_analysis}), consistently reaching $\sim$3$\times$ higher than expert drift at L45+ in both Stage~1 and Stage~2 despite an overall reduction in magnitudes. This suggests that pretrained expert networks already possess sufficient capacity for reasoning, and the role of RL is to incrementally learn appropriate token routing. Second, the two stages employ complementary modification strategies: Stage~1 concentrates its representation drift in the upper layers where output decisions are made, while Stage~2 additionally modifies mid-layer representations (L20--L24), suggesting that refining performance on acoustically challenging boundary cases requires adjustments to intermediate feature processing. Together, these findings paint a coherent picture: RL training teaches the model to route tokens through different expert combinations in the upper layers, enabling structured reasoning to emerge without rewriting the foundational acoustic and linguistic knowledge encoded in expert parameters.

\section{Conclusion}
\label{sec:conclusion}

We present Audio-DeepThinker, a progressive two-stage RL framework that enables high-quality CoT reasoning to emerge in LALMs through pure RL exploration, without supervised reasoning fine-tuning. Our hybrid reasoning similarity reward provides fine-grained supervision over reasoning quality, while the progressive curriculum first establishes foundational reasoning patterns via pure RL exploration, then refines them on acoustically challenging boundary cases. Audio-DeepThinker achieves state-of-the-art results on MMAR (74.0\%), MMAU-test-mini (78.5\%), and MMSU (77.26\%), winning 1st Place in the Interspeech 2026 Audio Reasoning Challenge. Interpretability analyses further reveal that RL training concentrates its modifications in upper-layer MoE gating mechanisms, providing mechanistic insights into how audio reasoning capabilities emerge through exploration. This finding also points to a promising direction for future work: parameter-efficient RL strategies that freeze expert parameters and optimize gating networks alone, potentially reducing training costs while preserving reasoning quality.

\bibliographystyle{IEEEtran}
\bibliography{mybib}

\appendix
\onecolumn
\section{Prompt Templates}
\label{sec:prompts}

This appendix provides the detailed prompt templates used in our framework.

\subsection{CoT Generation Prompt}
\label{ssec:cot_prompt}

The following prompt is used to generate ground truth chain-of-thought reasoning for training data construction using DeepSeek V3.1:

\begin{tcolorbox}[
    colback=gray!10,
    colframe=gray!50,
    title={\textbf{CoT Generation Prompt}},
    fonttitle=\bfseries,
    breakable,
    enhanced
]
\small
\texttt{Provide reasoning for why "\{answer\}" is the correct answer.}\\[0.5em]
\texttt{Audio Description:}\\
\texttt{\{qwen3\_captioner\}}\\[0.5em]
\texttt{Question: \{question\}}\\
\texttt{Options: \{options\}}\\
\texttt{Answer: \{answer\}}\\[0.5em]
\texttt{IMPORTANT:}\\
\texttt{1. If audio Description is completely unrelated to the question and answer $\rightarrow$ output only "None"}\\
\texttt{2. Otherwise, write 200-600 word reasoning that:}\\
\texttt{\quad - Identifies key audio evidence}\\
\texttt{\quad - Explains how it points to "\{answer\}"}\\
\texttt{\quad - Rules out other options}\\
\texttt{\quad - Sounds like analysis, not pre-knowledge}\\
\texttt{3. NO meta-commentary, NO repetition}\\
\texttt{4. Start immediately with your analysis}\\[0.5em]
\texttt{Analysis:}
\end{tcolorbox}

\subsection{Reasoning Similarity Evaluation Prompt}
\label{ssec:sim_prompt}

The following prompt is used by the LLM evaluator (Qwen3-235B-A22B-Instruct) to assess the similarity between generated reasoning and the reference CoT:

\begin{tcolorbox}[
    colback=blue!5,
    colframe=blue!40,
    title={\textbf{Reasoning Similarity Evaluation Prompt}},
    fonttitle=\bfseries,
    breakable,
    enhanced
]
\small
\texttt{You are an expert in evaluating reasoning similarity. Compare the following two reasoning processes and rate their similarity on a 0.0-1.0 scale (0.1 increments).}\\[0.5em]
\texttt{Consider:}\\
\texttt{1. Do they follow similar logical paths?}\\
\texttt{2. Do they cover the same key steps?}\\
\texttt{3. Are the reasoning strategies aligned?}\\
\texttt{4. Is the depth of analysis comparable?}\\[0.5em]
\texttt{Reference Reasoning (Ground Truth):}\\
\texttt{\{gt\_think\}}\\[0.5em]
\texttt{Generated Reasoning:}\\
\texttt{\{pred\_think\}}\\[0.5em]
\texttt{Output ONLY a number between 0.0 and 1.0. No explanation.}
\end{tcolorbox}

\subsection{Model System Prompts}
\label{ssec:system_prompts}

The following system prompts are used for the Qwen3-Omni-30B-A3B-Instruct and Qwen3-Omni-30B-A3B-Thinking models during evaluation:

\begin{tcolorbox}[
    colback=green!5,
    colframe=green!40,
    title={\textbf{Qwen3-Omni-Instruct System Prompt}},
    fonttitle=\bfseries,
    breakable,
    enhanced
]
\small
\texttt{You are an audio understanding model that answers multiple choice questions based on audio content. You MUST always start your reasoning with <reasoning> and end with </reasoning> tags, followed by the exact option text, i.e., <reasoning>your detailed reasoning process</reasoning>}\\
\texttt{selected option text}
\end{tcolorbox}

\begin{tcolorbox}[
    colback=orange!5,
    colframe=orange!40,
    title={\textbf{Qwen3-Omni-Thinking System Prompt}},
    fonttitle=\bfseries,
    breakable,
    enhanced
]
\small
\texttt{You are an audio understanding model that answers multiple choice questions based on audio content.}
\end{tcolorbox}

\section{QA Example with Caption and CoT}
\label{ssec:qa_example}

The following example illustrates a complete QA sample including the audio caption and chain-of-thought reasoning:

\begin{tcolorbox}[
    colback=white,
    colframe=gray!50,
    title={\textbf{Sample QA with Caption and CoT}},
    fonttitle=\bfseries,
    breakable,
    enhanced
]
\small
\textbf{Query:}\\
\texttt{<audio>What specific environmental sound is clearly audible beneath the conversation? Please choose the answer from the following options: [`Doors opening repeatedly', `Wind blowing gently', `Footsteps crunching steadily', `Traffic passing by'].}\\[0.5em]

\textbf{Solution:}\\
\texttt{<answer>Footsteps crunching steadily</answer>}\\[0.5em]

\textbf{Qwen3-Captioner Output:}\\
\texttt{The audio clip opens in a large, reverberant indoor environment, likely a Polish train station concourse or terminal. A female voice, close to the microphone and speaking in Standard Polish, calmly explains, ``U nas ju\.z trzyna\'scie, ale... wybieramy jego spokojn\k{a}, a dzisiaj...,'' The background is filled with a constant murmur of voices from a crowd of adults. The acoustic space is characterized by hard surfaces, creating noticeable echo and a sense of spaciousness. Throughout the recording, there is a persistent low-frequency mechanical hum and a rhythmic metallic clatter, likely from the movement of a train on tracks.}\\[0.3em]

\textbf{Chain-of-Thought Reasoning:}\\
\texttt{The audio description provides no direct evidence of footsteps, making this a process of elimination based on the environmental and acoustic clues present. The option ``Doors opening repeatedly'' is ruled out by the nature of the metallic sounds described. A single, distinct ``sharp metallic click'' is heard, suggestive of a one-time mechanical action like a ticket gate. The term ``repeatedly'' is not supported. ``Wind blowing gently'' is thoroughly inconsistent with the entire setting. The audio is explicitly set in a ``large, reverberant indoor environment.'' ``Traffic passing by'' is also incongruous with the indoor location. The mechanical sounds present are specifically identified as originating from within the transit environment. This process of elimination leaves ``Footsteps crunching steadily'' as the only logically plausible option. A busy train station concourse with hard surfaces would inherently have the sound of people walking as a fundamental layer of its ambient bed.}
\end{tcolorbox}

\newpage
\section{Training Hyperparameters}
\label{sec:hyperparameters}

\begin{table}[htbp]
\centering
\caption{Training hyperparameters for GDPO-based RL fine-tuning of Qwen3-Omni-30B-A3B-Instruct.}
\label{tab:hyperparameters}
\resizebox{0.77\columnwidth}{!}{
\begin{tabular}{ll}
\toprule
\textbf{Hyperparameter} & \textbf{Value} \\
\midrule
\multicolumn{2}{l}{\textit{Model \& Parallelism}} \\
Base Model & Qwen3-Omni-30B-A3B-Instruct \\
Tensor Model Parallel Size & 4 \\
Expert Model Parallel Size & 4 \\
Pipeline Model Parallel Size & 2 \\
Context Parallel Size & 1 \\
\midrule
\multicolumn{2}{l}{\textit{GDPO Training}} \\
Loss Type & GRPO \\
$\beta$ (KL penalty coefficient) & 0.001 \\
Importance Sampling Level & Sequence \\
Number of Generations per Prompt & 8 \\
\midrule
\multicolumn{2}{l}{\textit{Optimization}} \\
Learning Rate & $1 \times 10^{-6}$ \\
LR Warmup Fraction & 0.01 \\
Global Batch Size & 224 \\
Micro Batch Size & 4 \\
Steps per Generation & 4 \\
Max Epochs & 1 \\
Numerical Precision & BF16 \\
Train Type & Full Fine-tuning \\
\midrule
\multicolumn{2}{l}{\textit{Sequence \& Generation}} \\
Max Input Length (tokens) & 4096 \\
Max Completion Length (tokens) & 1024 \\
Sampling Temperature & 1.0 \\
Top-$p$ (nucleus sampling) & 0.99 \\
Top-$k$ & 50 \\
Overlong Filter & \cmark \\
Dynamic Sampling & \xmark \\
\midrule
\multicolumn{2}{l}{\textit{Reward Functions}} \\
\texttt{external\_base\_reward} & weight = 1 \\
\texttt{external\_consistency\_reward} & weight = 1 \\
\texttt{external\_hybridsimilarity\_reward} & weight = 1 \\
\midrule
\multicolumn{2}{l}{\textit{Saving \& Logging}} \\
Checkpoint Save Interval & every 100 steps \\
Log Interval & every 10 steps \\
Dataloader Workers & 8 \\
\bottomrule
\end{tabular}}
\end{table}

\end{document}